\documentclass[showpacs,twocolumn]{revtex4}

\begin{document}

\title{Reply to Comments on ``Invariance of the tunneling method for
  dynamical black holes''} 

\author{S. A. Hayward$^a$,
R. Di Criscienzo$^b$, M. Nadalini$^b$, L. Vanzo$^b$, S. Zerbini$^b$}

\address{$^a$Center for Astrophysics, Shanghai Normal University, 100
Guilin Road, Shanghai 200234, China}

\address{$^b$Dipartimento di Fisica, Universit\`a di Trento and INFN,
Gruppo Collegato di Trento, Italia}


\date{14th September 2009}

\begin{abstract}
We point out basic misunderstandings about quantum field theory and general 
relativity in the above Comments. In reply to a second comment on our first 
reply by the same author, we also identify precisely where the author's 
original calculation goes wrong and correct it, yielding the same local Hawking 
temperature as obtained by the Hamilton-Jacobi method.   
\end{abstract}
\maketitle

In observance to the ArXiv policy about repeated submissions on the same 
subject, we attach (and replace) to a previous submission  further 
considerations in reply to a second comment (ArXiv:0909.3800) by the same 
author, which for some reason was not subjected to the same policy. For the 
sake of honesty, we repeat exact text of our first reply in the first section, 
postponing to a second section the new considerations referring to the second 
comment. 

\section{First reply}
Hawking radiation from black holes \cite{Haw} is a well established
prediction of quantum field theory on curved space-times, confirmed
by multiple independent methods, see for example
\cite{dewitt,BD,wald,fulling,visser,igor}.  
We have recently generalized this prediction from static to dynamic
black holes 
\cite{CNVZZ,HCNVZ,CHNVZ}. Conversely, an article of
Pizzi \cite{Piz1} claims that Hawking radiation is a myth which has
fooled almost everyone but himself. For some reason, he has chosen
to write it in the form of a Comment on our article \cite{CHNVZ}.
While we feel no need to defend anything in our article, here we
point out briefly the main errors in the reasoning in \cite{Piz1}.

1. The author states that ``the action\dots along the classical light-like ray
is\dots constant'' and therefore ``no imaginary part in the action can
appear''. However, Hawking radiation is not a prediction of classical physics
but of quantum field theory. In the WKB approximation, the action is not
constant, but indeed rapidly varying.

2. The author states that an ``infinitesimally small neighbourhood of the
horizon\dots can be covered by Minkowski coordinates''. This is incorrect. The
correct statement is that connection coefficients, being first derivatives of
the metric, can be set to zero at a point. Surface gravity and the
corresponding temperature are curvature invariants, which involve second
derivatives of the metric and cannot be set to zero at a point.

3. There is a confusion of partial derivatives in the author's equation (3). He
appears to be solving the null geodesic equation rather than the
Hamilton-Jacobi equation.

4. The author's procedure for dealing with the pole in the action is
inequivalent to the standard one, namely the Feynman $i\epsilon$ procedure or
something equivalent, which corresponds to the desired physical boundary
conditions. It has not been justified and is used by no other author as far as
we are aware.

\section{Second reply}
1. The author states in \cite{Piz2} that ``to affirm that because of quantum 
theory we will have in the principal WKB approximation an action of
essentially  different character is nonsense''. This is untrue. When
going from classical  
physics to quantum field theory, qualitatively new features occur
which cannot  be found in the classical limit. 

2. The author has not answered our point 2, that a fundamental misunderstanding 
about general relativity was the basis of one of his two arguments. 

3. The author states in \cite{Piz2} that ``the equation (3)\dots is a trivial 
equation which in no way can contain any confusion''. Here we explain the 
confusion. The equation reads 
\begin{eqnarray}\label{I}
I&=&\int\left(\partial_rIdr+\partial_vIdv\right)\\\label{I2}
&=&\int\left(\partial_rI-\frac12\frac{dv}{dr}e^\Psi C\partial_rI\right)dr
\end{eqnarray}
where
\begin{equation}
\frac{dv}{dr}=\frac{2e^{-\Psi}}C
\end{equation}
was obtained by solving the null geodesic equation. This derivative is
infinite  
at the horizon $C=0$, so the above manipulation is invalid. Also 
\begin{equation}\label{k}
\partial_r I=-\frac{2e^{-\Psi}\partial_vI}C
\end{equation}
is infinite at the horizon, as the angular frequency 
\begin{equation}\label{omega}
\omega=e^{-\Psi}\partial_vI
\end{equation}
is finite. So (\ref{I2}) is meaningless, infinity minus infinity. 

Directly substituting (\ref{k}), (\ref{omega}) into (\ref{I}) yields 
\begin{equation}
I=\int e^\Psi\omega dv-\int\frac{2\omega}C dr
\end{equation}
which is the same expression as obtained by the Hamilton-Jacobi method 
\cite{HCNVZ,CHNVZ}. The first term is finite, while the second has a pole at 
$C=0$. This is the famous pole in the action. 

4. Therefore the author's claim in \cite{Piz2} that there ``are no
poles in the  
integrand of the action'' is false. 

5. The author suggests in \cite{Piz2} ``to focus firstly on the
simpler case of  
the Schwarzschild black hole, where only very well known formulas are used and 
any eventual mistake is easy to be discovered''. Indeed, in this case it is 
easy to see that $dr/dv$ vanishes along the horizon, where $r$ is constant.

\medskip
SAH was supported by the National Natural Science Foundation of China under
grants 10375081, 10473007 and 10771140, by Shanghai Municipal Education
Commission under grant 06DZ111, and by Shanghai Normal University under grant
PL609.

\end{document}